\documentclass[10pt,journal,compsoc]{IEEEtran}
\ifCLASSINFOpdf
\else
\fi
\begin{document}
%
\title{A Privacy Preserved and Cost Efficient Control Scheme for Coronavirus Outbreak Using Call Data Record and Contact Tracing}
%
%
%
%

\author{Shibli Nisar, 
        Syed Muhammad Ali Zuhaib, Abasin Ulasyar and Muhammad Tariq
\IEEEcompsocitemizethanks{\IEEEcompsocthanksitem S.Nisar, M. Zuhaib and A. Ulasyar are with the Department
of Electrical Engineering, National University of Sciences and Technology-NUST, Pakistan.\protect\\
E-mail: shiblinisar@mcs.edu.pk
\IEEEcompsocthanksitem M. Tariq is with Electrical Engineering Department, National University of Computer and Emerging Sciences, Pakistan. He is also a visiting research collaborator at Princeton University, NJ, USA. Email: mtariq@princeton.edu}
\thanks{A PREPRINT Oct, 2020.}}

%
%

\markboth{A PREPRINT, October~2020}%
{Shell \MakeLowercase{\textit{et al.}}: Bare Advanced Demo of IEEEtran.cls for IEEE Computer Society Journals}
%



\IEEEtitleabstractindextext{%
\begin{abstract}

Coronavirus or COVID-19, which has been declared pandemic by the World Health Organization, has incurred huge losses to the lives of people throughout the world. Although, the scientists, researchers and doctors are working round the clock to develop a vaccine for COVID-19, it may take a year or two to make a safe and effective vaccine available for the world. In current circumstances, a solution must be developed to control or stop the spread of the virus. For this purpose, a novel technique based on call data record analysis (CDRA)and contact tracing is proposed that can effectively control the coronavirus outbreak. A positive coronavirus patient can be traced through CDRA and contact tracing. The technique can track the path traversed by the patient and collect the cell numbers of all those people who have met with the patient. Keeping in tact the privacy of this group of people, who are contacted through their cell numbers so that they can isolate themselves till the result of their coronavirus test arrives. If a test result of a person comes positive among the group, then he/she must be isolated and same CDRA and contact tracing procedures are adopted for that person. A COVID-19 patient is geo tagged and alerts are sent if any violation of isolation is done by the patient. Moreover, the general public is informed in advance to avoid the path followed by the patients. This cost effective mechanism is not only capable to control the coronavirus outbreak but also helps in isolating the patient in his/her house. 
\end{abstract}

}

\maketitle

\IEEEdisplaynontitleabstractindextext

%
\IEEEpeerreviewmaketitle

\section{Introduction}
\label{sec:introduction}
Coronavirus disease (COVID-19) is an infectious disease, which is caused by a newly discovered coronavirus \cite{a1}. So far, the disease has spread like a wildfire and almost reached to the entire world by affecting millions of people \cite{s11}. The most scary side of the disease is that, it is a novel virus accompanied with unknown myths and realities \cite{a3}. Covid-19  belongs to a family of ribonucleic acid (RNA), which is basically a virus family named as coronaviridae \cite{a4}. To make the curve of coronavirus outbreak flatten, the governments have imposed lockdown in their countries, enforced border shutdowns and travel restrictions. Consequently, many social and economic sectors are affected such as petroleum and oil industry \cite{a5}, agriculture \cite{a77}, manufacturing industry \cite{a8}, education \cite{a9}, finance industry \cite{a10}, healthcare and pharmaceutical industry \cite{a11}, tourism industry \cite{a12}, real estate and housing sector \cite{a133}, sport sector \cite{a13}, information technology and media industry \cite{f11}. Moreover, social distancing and lockdown measures to control the spread of virus have caused severe impacts on social life in terms of of domestic violence, that includes emotional, physical and sexual abuse \cite{s22}.  

With the emergence of coronavirus, researchers from all over the world are actively engaged to find the drug or vaccine and discover different ways to control its spread. Since, there is no vaccine or drug available to cure this deadly disease, therefore, the best solution is to control the wide spread of the disease by social distancing and isolation \cite{p2}. Currently, the main concern is to minimize the coronavirus outbreak. In order to control the spread of the virus, this research work proposes a novel solution that can track, trace and terminate coronavirus disease by using call data record analysis (CDRA), which is commonly known as call detailed record. A call data record is a log file maintained by the telephone or mobile operators. The log file contains data such as, duration of calls made or received at a particular time, source and destination numbers, call type, international mobile equipment identity (IMEI), cell site detail, and latitude and longitude of base transceiver station (BTS) \cite{q1}. 

Almost all law enforcement agencies (LEA) use call data record (CDR) for tracking, tracing and investigation of criminals \cite{z1}. This is one of the most widely used techniques to cope with the criminals and have shown  very promising results \cite{z5}. LEAs use CDR analysis to track and trace the entire criminal network and find out the hidden connections between criminals. This paper proposes a technique that assists the governments to track and trace COVID-19 patients with the help of CDRA in the same manner that is usually done for criminals. The concern department make it sure to keep the privacy of a patent in tact. The technique not only identifies the  suspected cases through call data record of a patient but also it tracks the path followed by the COVID-19 patient.

\begin{figure}
\centering
\includegraphics[scale=1]{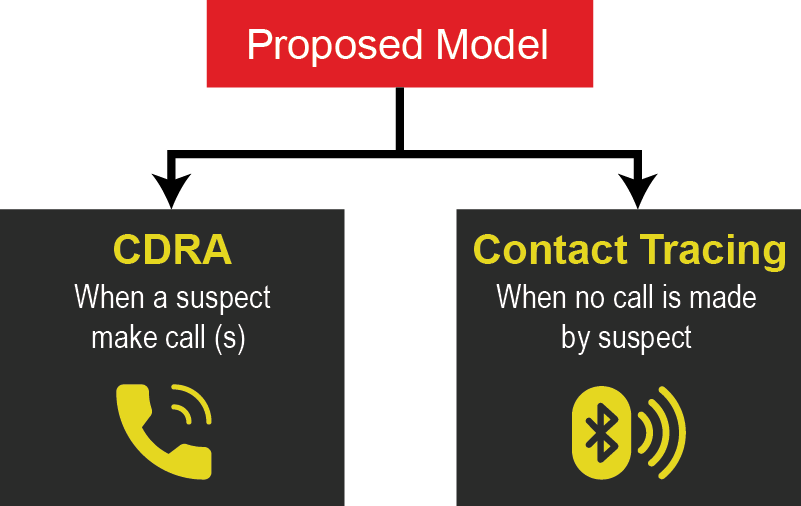}
\caption{Block diagram of the proposed model}
\label{wwf7}
\end{figure}

One of the main limitations of the CDRA is when a suspect goes into the crowd without making a phone call. In order to overcome this limitation contact tracing based on exposure notification system is used in parallel to track the path traversed made by the patient without call history \cite{shibli11}. The exposure notifications system (ENS), originally known as the privacy-preserving contact tracing project, is a framework and specification developed by Apple and Google to facilitate digital contact tracing during the COVID-19 pandemic \cite{shibli44}. Hence the proposed model is capable to track and trace COVID-19 patients with and without call history as shown in Figure \ref{wwf7}.

Rest of the paper is organized as follow: Section \ref{CDRM-22} presents the CDMA framework. Section \ref{contact-tracing} discusses contact tracing framework. Section \ref{conclusion2} presents the conclusive remarks, potential outcomes of the proposed model, and some future directions.

\section{CDRA Framework}
\label{CDRM-22}
\begin{figure}
\centering
\includegraphics[scale=.5]{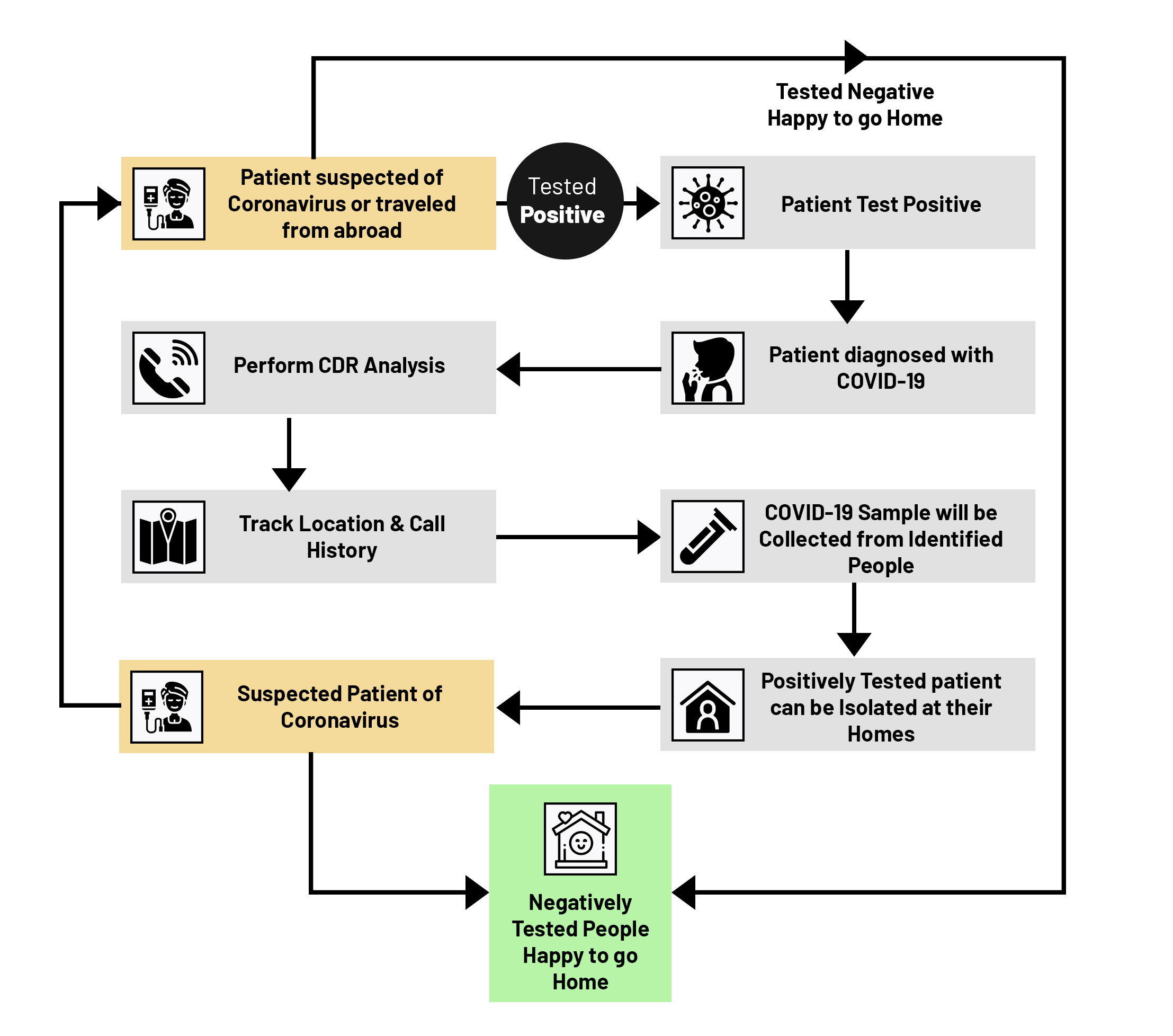} 
\caption{Complete flowchart of the proposed CDRA model}
\label{12a}
\end{figure}

In order to curb the spread of COVID-19, a model is required that can immediately locate the coronavirus patients by using the geo location. This type of location can be found with the help of tower signals and call history of a patient. If we track, control and isolate, the spread can be controlled in a few weeks. Otherwise, if we ignore instruction at the start, it will be an immense damage. This can be easily done through the CDRA. Through CDRA, one can easily track the location and call history of a patient and can reach all possible suspects who met the patient. By doing so, one can easily stop or minimize the coronavirus outbreak. In addition, the potential possible carrier should be geo tagged and tracked by sending alerts if he or she violates quarantine or self-isolation. This will help the people to isolate themselves and not becoming the source of coronavirus spread. Detailed flow diagram of the proposed model is shown in Figure \ref{12a}

A CDR provides metadata. Metadata is a set of data that describes and gives information about other data. Such as how a specific phone number and/or user is utilizing the phone system. This metadata includes different information about the call but more relevant information it contained is summarized in Table \ref{t11}.

\begin{table}

\begin{center}
\caption{Relevant attributes of CDR}
\label{t11}
\begin{tabular}[width=\textwidth]{ |l| } 
\hline
\textbf{Call Data Record}  \\
\hline
When the call took place (date and time)   \\
    \hline

   How long the call lasted (in minutes)   \\
    \hline
    
    Who called whom (source and destination phone numbers) \\
    \hline
    
 What kind of call was made (inbound, outbound, toll-free) \\
    \hline
    
International Mobile Equipment Identity (IMEI) detail   \\
  \hline
    
   How much the call cost (based on a per minute rate)  \\
    \hline
    Information about Base Transceiver Station (Location)  \\
    \hline
    longitude and latitude of BTS when call is made/received   \\
    \hline
\end{tabular}
\end{center}

\end{table}

CDRs can also include short messaging services (SMS) messaging metadata and any other official communications transmission. However, the contents of the messages/calls are not revealed through the CDR, thus keeping the privacy in tact. The call detail record simply shows that the calls or messages took place, and measures basic call properties. Initially, one has to perform CDR analysis on all those people who had recently traveled from abroad and could be source for spreading coronavirus in other people or those whose results are declared positive. If the patient is diagnosed with coronavirus, the hospital authorities will inform the government authority to perform CDR analysis on the patient's active sim with the consent of patient. After CDRA of a patient, the concern government team will contact people with whom the patient might be interacted (these people are known to be suspected people). Ambulance service will go to suspected patients' home and will get sample for coronavirus. During the process, suspected people will be guided to stay at their homes or remain isolated until final result of the test is revealed. If the test comes positive, the patient will be sent to a hospital. Similar procedure will be repeated for all those people to whom the patient has been interacted and their corona test was positive. The potential outcomes of the proposed model are listed below:
\begin{itemize}

\item tracking of COVID-19 patients,

\item monitoring of isolated patients, 

\item tracking of suspected ones,

\item inform the mass about the safest path to use.

\end{itemize}

\subsection{Case Study 1}
In order to integrate the information and interpret the outcome of the proposed model, one complete case study based on fiction is discussed in this section. If the subject’s corona test comes positive then the very first step is to quarantine the subject and asked the mobile operator for his/her CDR. CDRA starts just after getting the CDR from the mobile operator. After receiving the CDR, it starts in the format given in Table \ref{t22}. The table contains the most relevant attributes of the CDR.


\begin{table*}
\begin{center}
\caption{Relevant information contained in patient’s CDR}
\label{t22}
\begin{tabular}{ | c| c | c | c | c | c | c | c |  } 
\hline
\textbf{Date \& Time} &\textbf{A party}  & \textbf{B party} & \textbf{Call Type} &\textbf{IMEI}  &\textbf{Cell Site}  & \textbf{Latitude} & \textbf{Longitude}  \\
\hline
08/01/2020 11:13:04 & 1234567890 & 12121212121& Call Outgoing & 3530030719058&Plot \# 1,
Rawalpindi

&33.52292 & 73.23864  \\
    \hline

  08/01/2020 12:30:02 & 1234567890& 13131313131& Call Incoming &3530030719058 & Plot \# 2,
Rawalpindi

& 33.5026& 73.1965  \\
    \hline
    
   05/01/2020 12:35:56 &1234567890 & 14141414141&Call Outgoing &3530030719058 &Plot \# 2,
Rawalpindi

 &33.5026 &73.1965 \\
    \hline
    
 05/02/2020 14:41:39 & 1234567890& 15151515151& Call Incoming&3530030719058 & Plot \# 1,
Islamabad
&33.65647 &73.0367 \\
    \hline
    
05/02/2020 14:55:42 &1234567890 & 17171717171&Call Incoming &3530030719058 &Plot \# 2,
Islamabad
 &33.65482 &73.04096  \\
  \hline
    
   05/03/2020 17:16:29  & 1234567890& 18181818181& Call Outgoing&3530030719058 &Plot \# 3,
Islamabad
 &33.66394 &73.07612 \\
    \hline
    05/03/2020 17:32:06 & 1234567890& 18181818181 &Call Outgoing & 3530030719058& Plot \# 4,
Islamabad
& 33.5995&73.13191  \\
    \hline
   05/04/2020 17:46:14  &  1234567890& 19191919191 &Call Incoming &3530030719058 & Plot \# 5,
Islamabad
& 33.5759& 73.1506 \\
   
    \hline

    05/05/2020 17:49:33  & 1234567890 &12121212121 & Call Incoming& 3530030719058& Plot \# 6,
Islamabad
 & 33.57096& 73.1452 \\
   
    \hline
    05/06/2020 17:51:23  & 1234567890 &14141414141 &Call Incoming &3530030719058 &Plot \# 6,
Islamabad
&33.57096 & 73.1452  \\
   
    \hline
    05/06/2020 17:52:35  & 1234567890 & 17171717171& Call Outgoing& 3530030719058& Plot \# 8,
Islamabad&33.569 &73.1391\\
   
    \hline
    05/07/2020 10:45:34  &  1234567890& 14141414141&Call Outgoing &3530030719058 & Plot \# 9,
Islamabad&33.65482 & 73.04096  \\
   
    \hline
    05/07/2020 12:39:40  &  1234567890& 16161616161& Call Incoming&3530030719058 & Plot \# 9,
Islamabad&33.65482 & 73.04096 \\
   
    \hline
    05/07/2020 12:47:11 &  1234567890& 15151515151 &Call Incoming &3530030719058 & Plot \# 10,
Islamabad& 33.66482& 73.04196  \\
   
    \hline
    05/07/2020 14:35:37  &1234567890  & 13131313131&Call Outgoing &3530030719058 & Plot \# 10,
Islamabad & 33.66482& 73.04196 \\
   
    \hline
\end{tabular}
\end{center}
\end{table*}

\begin{figure}
\centering
\includegraphics[scale=1.1]{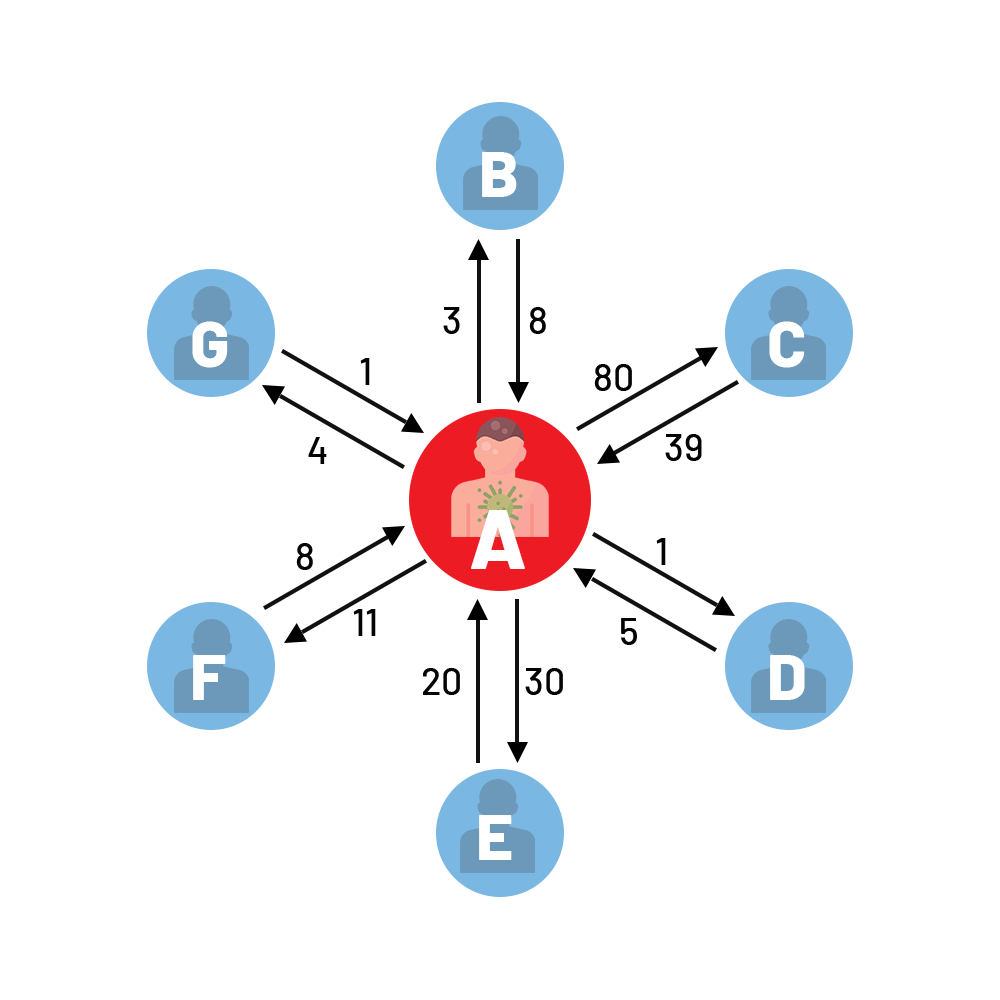}
\caption{Graphical interpretation of patient’s CDR}
\label{Cqq}
\end{figure}

CDR of the subject’s contained all the relevant information through which one can easily track and trace the patient through his/her cellular contacts along with path followed by that subject. From CDR, we can get the information about the calls made and received by the patient along with date and time. CDR contains the duration of all calls made and received. CDR includes information about the BTS with which the subject is connected. CDR also holds the latitude and longitude of BTS when a call is being made or received. Before starting the CDR analysis the very first step is to remove the past irrelevant data. This can be easily done by deleting all past entries and keeping the data of few couple of weeks. Doing so, one can easily limit the CDR data in the desired dates. This will help to analyze the small amount of data efficiently and accurately. In order to clearly visualize the data and extract hidden patterns from the CDR, it is recommended to convert tabular data of CDR into graphical one. Different tools are available to convert tabular data into graphical one. One such tool is IBM I2 analyze. This tool will convert the tabular data  into graphical one as shown in Figure \ref{Cqq}.

\begin{figure}
\centering
\includegraphics[scale=.35]{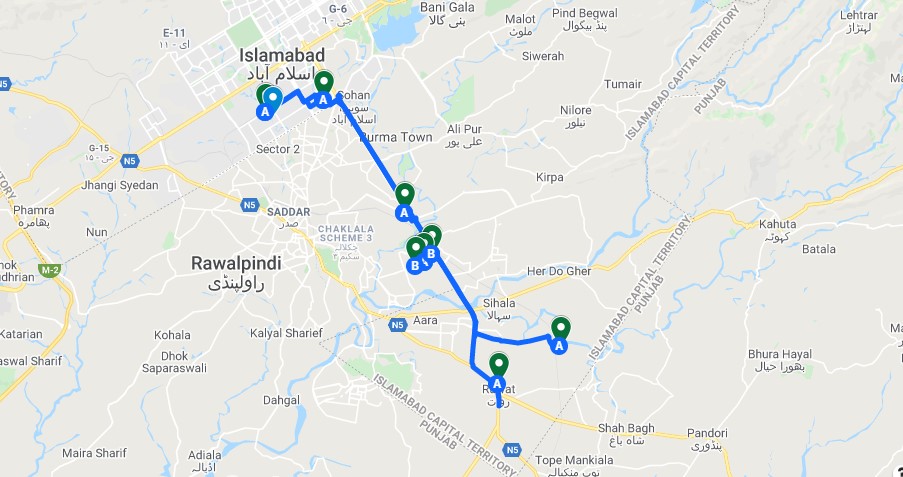}
\caption{Path followed by Patient (node A)}
\label{dd}
\end{figure}

Nodes in Figure \ref{Cqq} represent the calling parties. While the frequency of calls are shown through the edges. Node A represents the subject, while nodes B, C, D, E, F and G are the one with whom the patient was in contact. From Figure \ref{Cqq} one can easily interpret that the subject is frequently in contact with party C, with a total of 80 outgoing calls and 39 incoming calls. Patient is in least contact with party G with only 4 outgoing calls and 1 incoming call. While investigating a COVID-19 patient, Figure \ref{Cqq} depicts a vital role to memorize the patient with whom he/she has met. CDR analysis outcomes will help the patient to reduce the chances to miss anyone with whom the patient has met. In fact all the parties those were in contact with the patient will be contacted. It will be confirmed whether they have met the patient or not. Through CDR of a patient, one can easily improve the accuracy of tracking and tracing the suspect ones. After identifying the people who have met with the patient will be sent into the quarantine. During the process the suspected people will be guided to stay at their homes or remain isolated until the final result of the test is declared. If the test result is positive then the patient will be sent to hospital. Similar procedure will be adapted to all those people to whom the patient has been interacted. In addition to this through CDRA one can easily track the path followed by the patient using cell site detail and latitude and longitude. With the help of cell site data and latitude and longitude of cell site available in Table \ref{t22}, one can easily plot the map of the COVID-19 patient using Google map. Figure \ref{dd} shows the Google map plot with the help of cell site detail given in CDR. This map will be made available to general public and alerts will be generated to avoid this path.

\begin{figure}
\centering
\includegraphics[scale=1.1]{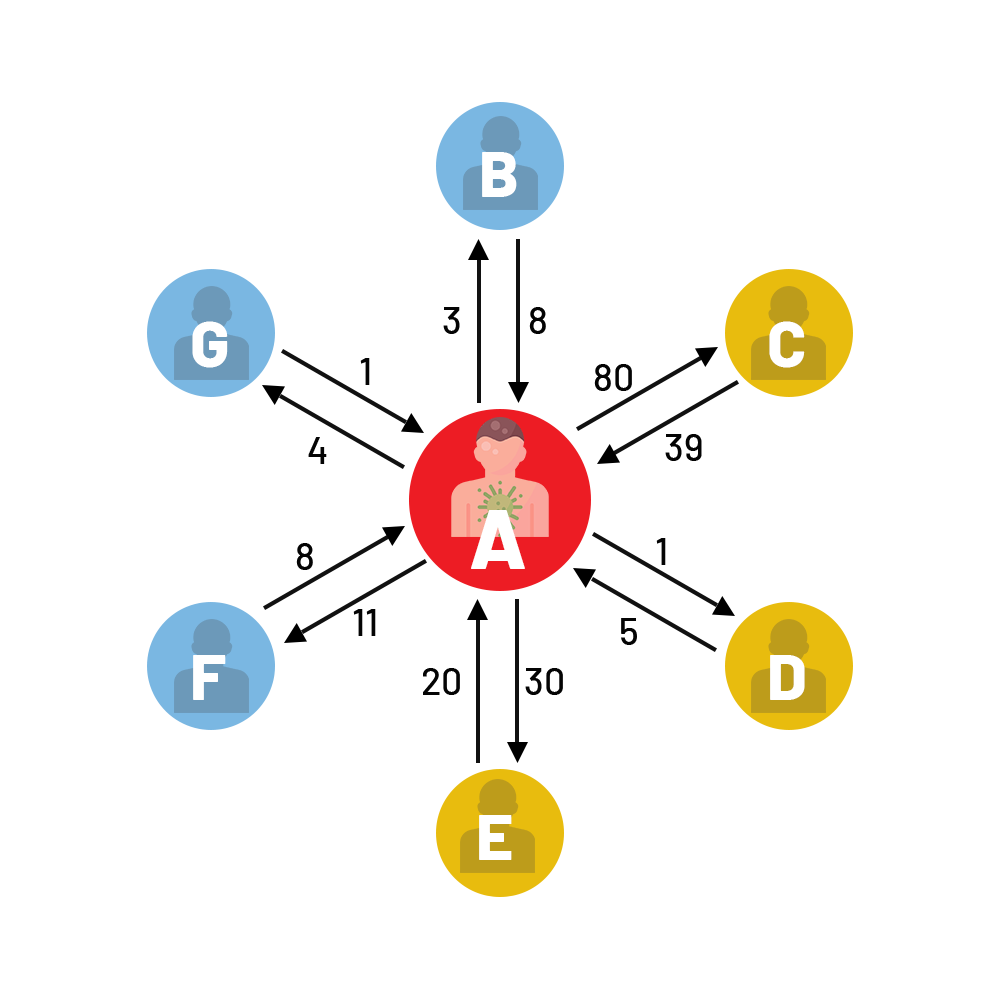}
\caption{Identification of coronavirus suspects}
\label{ww}
\end{figure}

After investigating patient A in the light of CDR graphical links. Let suppose patient A admits that during this period he/she has met with C, D and E. Therefore node C, D and E will be labeled as suspects as shown in Figure \ref{ww}. The concerned government department will immediately contact C, D and E and will ask them to quarantine themselves until and unless their corona tests come negative.

\begin{figure}
\centering
\includegraphics[scale=1.1]{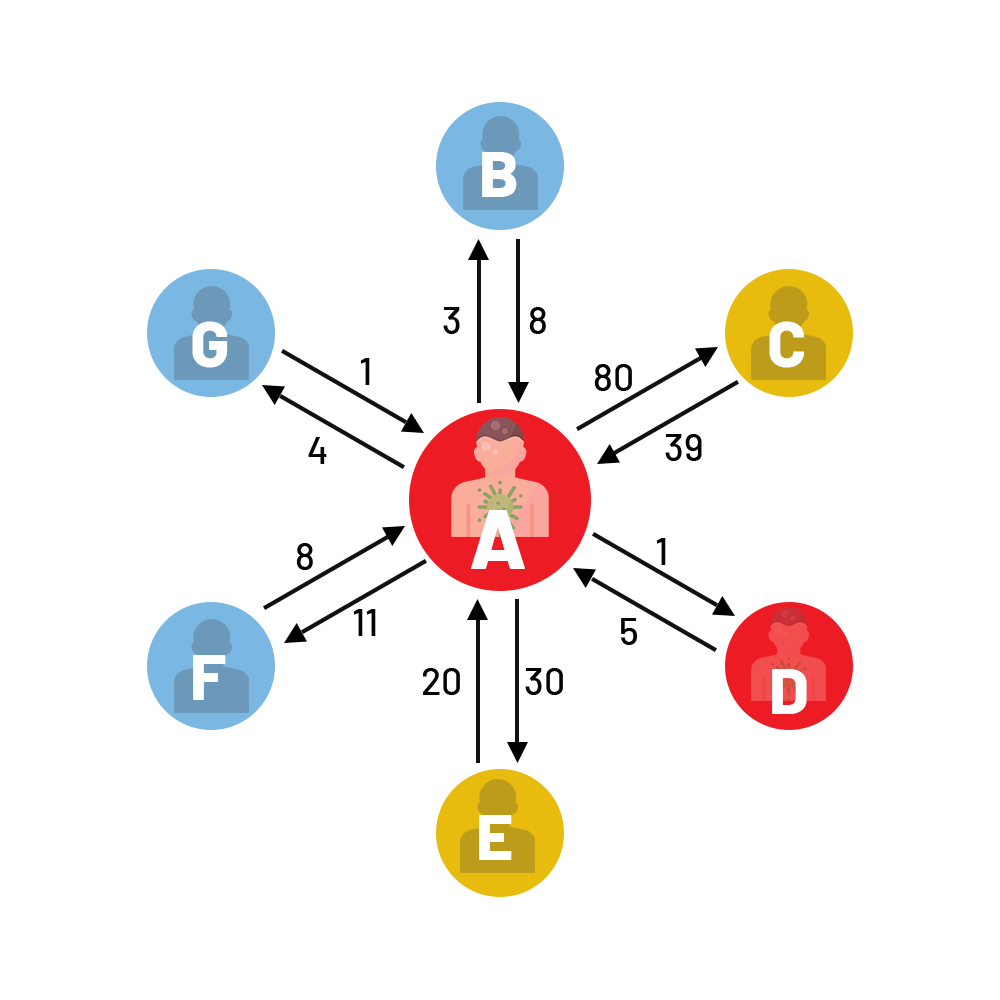}
\caption{Identification of coronavirus patient}
\label{ww1}
\end{figure}

Suppose that after testing the suspect ones, only D comes positive then this node will be labeled as patient and the graph will be modified as shown in Figure \ref{ww1}.

\begin{figure}
\centering
\includegraphics[scale=.7]{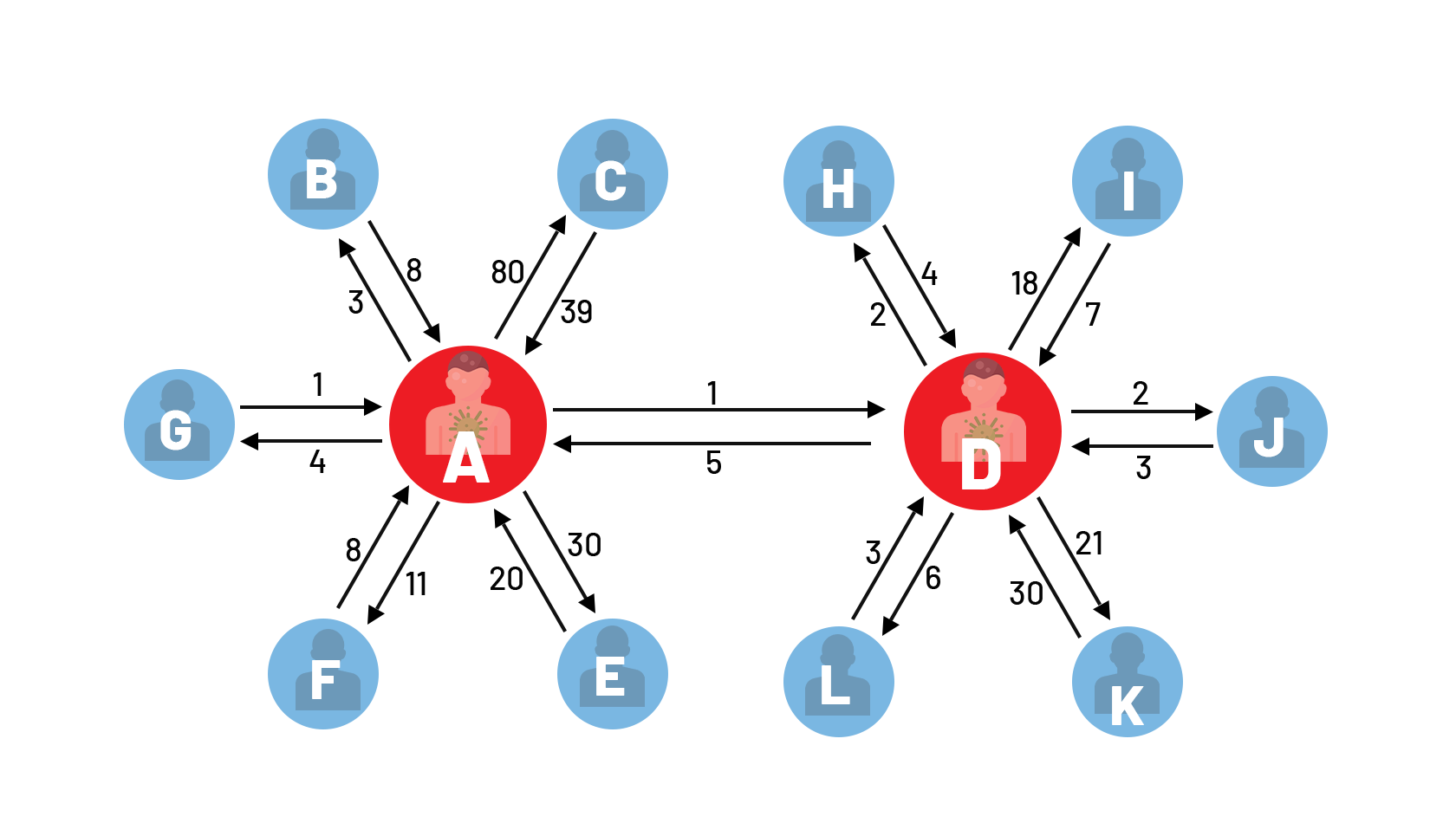}
\caption{Graphical interpretation of node D CDR}
\label{the22}
\end{figure}

Now the entire process will be repeated for node D i.e., CDRA of node D as shown in Table \ref{the1}. Its graphical interpretation is shown in Figure \ref{the22}.

\begin{table*}
\begin{center}
\caption{Relevant information contained in patient’s CDR (node D)}
\label{the1}
\begin{tabular}{ | c | c | c | c | c | c | c | c |  } 
\hline
\textbf{Date \& Time} &\textbf{A party}  & \textbf{B party} & \textbf{Call Type} &\textbf{IMEI}  &\textbf{Cell Site}  & \textbf{Latitude} & \textbf{Longitude}  \\
\hline
05/01/2020 11:13:04 & 9876543210 & 21121212121& Call Outgoing & 2110030719058&Plot \# 1,
Rawalpindi

&33.66822 & 73.23864  \\
    \hline

  05/08/2020 12:30:02 & 9876543210& 31131313131& Call Incoming &2110030719058& Plot \# 2,
Islamabad

& 33.66822& 73.23864  \\
    \hline
    
   05/08/2020 12:35:56 &9876543210 & 41141414141&Call Outgoing &2110030719058 &Plot \# 3,
Islamabad

 &33.67336 &73.01329 \\
    \hline
    
 05/09/2020 14:41:39 & 9876543210& 51151515151& Call Incoming&2110030719058 & Plot \# 4,
Islamabad
&33.68051 &73.01681 \\
    \hline
    
05/09/2020 14:55:42 &9876543210 & 71171717171&Call Incoming &2110030719058 &Plot \# 5,
Islamabad
 &33.6844 &73.01676  \\
  \hline
    
   05/10/2020 17:16:29  & 9876543210& 81181818181& Call Outgoing&2110030719058 &Plot \# 6,
Islamabad
 &33.6844  &72.98836 \\
    \hline
    05/11/2020 17:32:06 & 9876543210& 81181818181 &Call Outgoing & 2110030719058& Plot \# 6,
Islamabad
& 33.6844 &72.98836  \\
    \hline
   05/12/2020 17:46:14  &  9876543210& 91191919191 &Call Incoming &2110030719058 & Plot \# 6,
Islamabad
& 33.6844 & 72.98836 \\
   
    \hline

    05/12/2020 17:49:33  & 9876543210 &21121212121 & Call Incoming& 2110030719058& Plot \# 7,
Islamabad
 &33.6344 & 72.98846 \\
   
    \hline
    05/13/2020 17:51:23  & 9876543210 &41141414141 &Call Incoming &2110030719058 &Plot \# 7,
Islamabad
&33.6855  & 72.97736  \\
   
    \hline
    05/14/2020 17:52:35  & 9876543210 & 71171717171& Call Outgoing& 2110030719058& Plot \# 6,
Islamabad&33.6844  &72.988361\\
   
    \hline
    05/15/2020 10:45:34  &  9876543210& 41141414141&Call Outgoing &2110030719058 & Plot \# 8,
Islamabad&33.5544  & 72.96636  \\
   
    \hline
    05/15/2020 12:39:40  &  9876543210& 61161616161& Call Incoming&2110030719058 & Plot \# 10,
Islamabad&33.56844  & 72.98446 \\
   
    \hline
    05/16/2020 12:47:11 &  9876543210& 51151515151 &Call Incoming &2110030719058 & Plot \# 6,
Islamabad& 33.6844 &72.98836 \\
   
    \hline
    05/16/2020 14:35:37  &9876543210  & 31131313131&Call Outgoing &2110030719058 & Plot \# 6,
Islamabad & 33.6844 & 72.98836 \\
   
    \hline
\end{tabular}
\end{center}
\end{table*}

\begin{figure}
\centering
\includegraphics[scale=0.35]{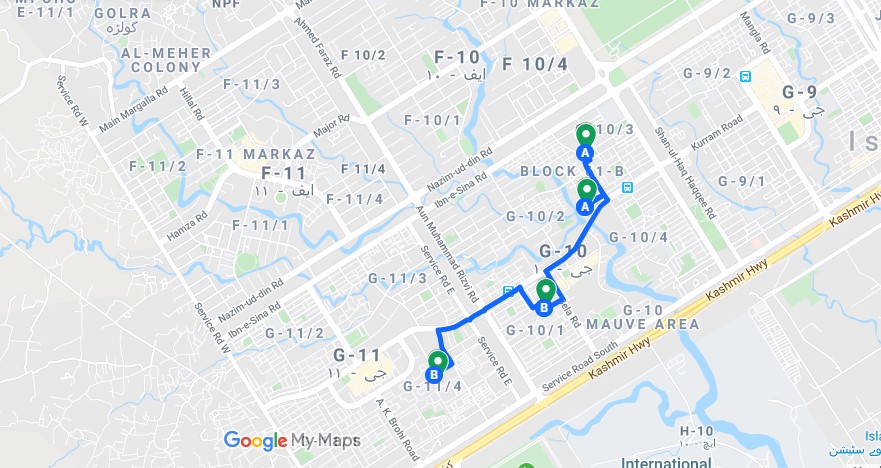}
\caption{Path followed by Patient (node D)}
\label{shibli111}
\end{figure}

Google map of the patient node D will be created with the help of data available in CDR of node D as shown in Figure \ref{shibli111}. The map will be made public with an alert to avoid the use of this path for couple of weeks. 

\begin{figure}
\centering
\includegraphics[scale=.7]{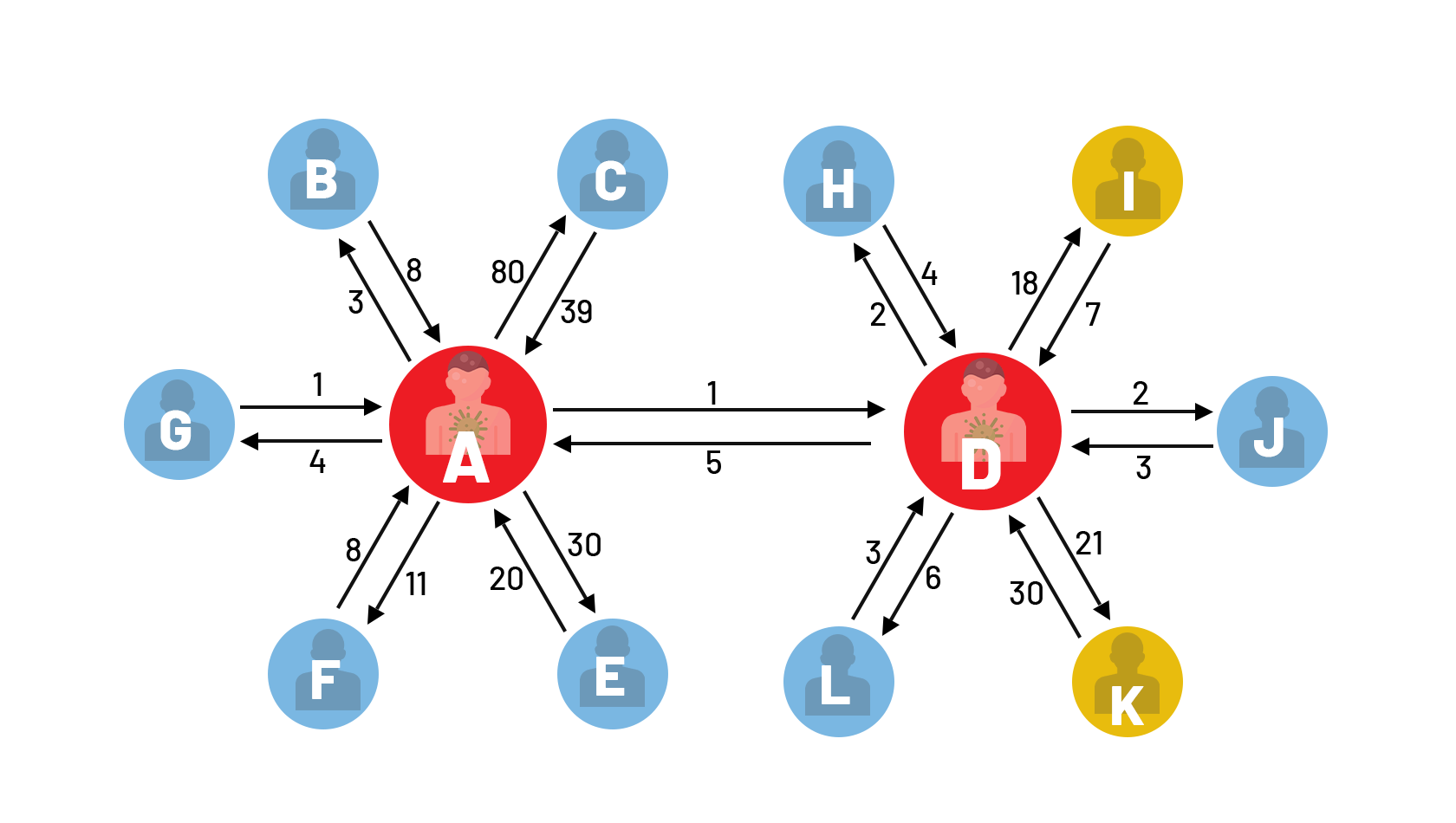}
\caption{Identification of coronavirus suspects}
\label{final1}
\end{figure}

After investigating node D, one can easily find out the suspects in the light of CDR graph linkages and path followed by the patient as shown in Figure \ref{the22}, \ref{shibli111}, respectively. Similarly, the same cycle will be repeated to dig out the suspected ones as shown in Figure \ref{final1}. This process will keep on repeating until the entire web of patients are traced. All COVID-19 patients will be geo tagged in order to track their quarantine and alerts will be generated if he or she violates isolation. 

\section{Contact Tracing Framework}
\label{contact-tracing}

Since the proposed model is based on CDR, which only works once a suspected person made a call and its log is maintained and tracked. However, in a scenario where a person who is infected, meet to another person without having a phone call or go into the crowd, then the model switches to exposure notifications framework. Exposure notifications system (ENS) previously known as contact tracing is a technique in which two nearby mobile devices exchange random sequence of keys and a log is maintained in a cloud storage to keep record of all the devices that have been remained in physical contact previously. This framework is developed by the joint efforts of Google and Apple. Most importantly this technique protect user’s privacy and security. 

\subsection{How it Works?}

\begin{figure}
\centering
\includegraphics[scale=1.2]{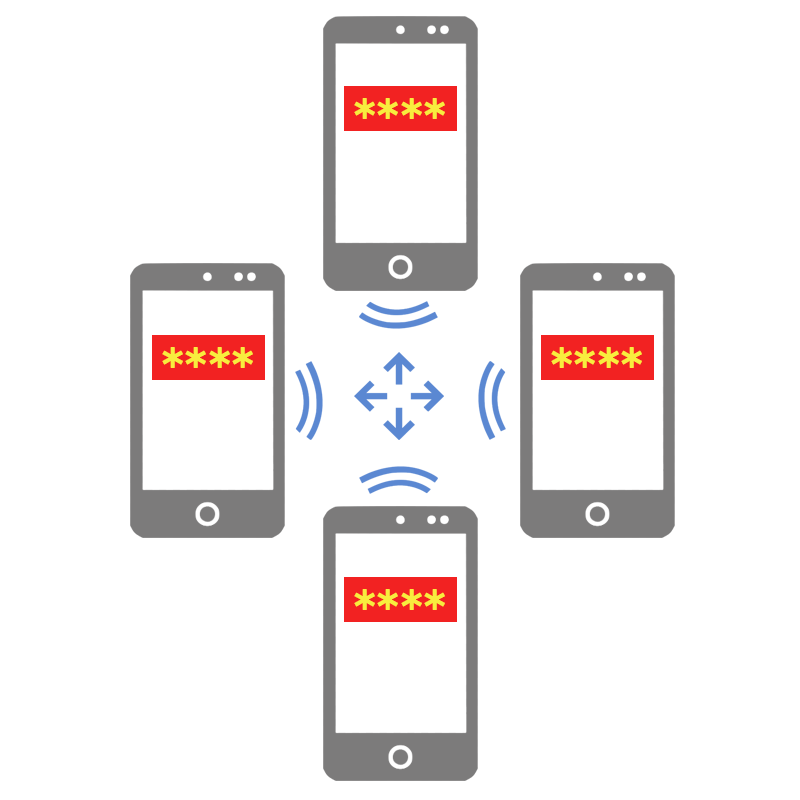}
\caption{Connectivity through Bluetooth}
\label{ddq}
\end{figure}

People all around the world continuously working on developing apps based on exposure notification system to help them in tracing the people that remained in contact with infected patients of Covid-19. Once an individual selects this model, a unique random key of 4 digits generates on that mobile device. This key is continuously changed after 10 to 20 minutes so that to ensure it cannot be used to recognize the location. A phone with ENS based app comes into contact with another device embedded with same app, both phones exchange these random keys with each other through bluetooth low energy (BLE) as shown in Figure \ref{ddq}. This not only ensures the privacy but also improves the  overall  power consumption. 

User’s phone periodically checks all the random keys that have been maintained to identify a patient tested positive with COVID-19. If there is a match exists, user and concern government department receive exposure notification on its phone. In this way one can overcome the limitation of CDR when a person comes into the contact with another without having a phone call.

\subsection{User Privacy}

Privacy always remains at first. In exposure notifications a user may allow it anytime and can disallow with its own consent. As discussed above, exposure notifications do not track location of your device and only works based on BLE. Exposure notification matching only occurs on the device and this framework does share identity with other people, Google, or Apple. Concerned authorities may ask for contact details but only with the consent of user.

\subsection{Case Study 2}
\begin{figure}
\centering
\includegraphics[scale=1.8]{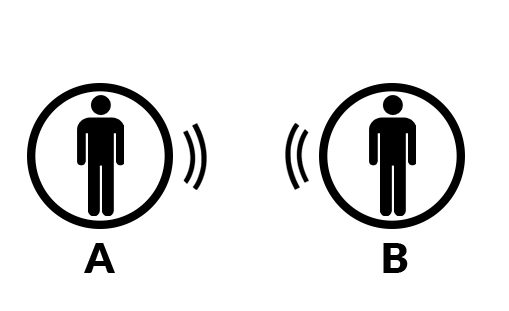}
\caption{Graphical interpretation of contact tracing }
\label{dde}
\end{figure}

In order to integrate the information and interpret the outcome of the proposed model, one complete case study based on hypothesis is discussed below. Two people named A and B go to a marketplace at the same time and place. They both have  contact tracing app installed. After the shopping, they wait for 10 minutes in check out queue. During this time interval their phones exchange random keys to each other and store information. Thus they know each other that they have remained in contact for a while as shown in Figure \ref{dde}.

\begin{figure}
\centering
\includegraphics[scale=1.8]{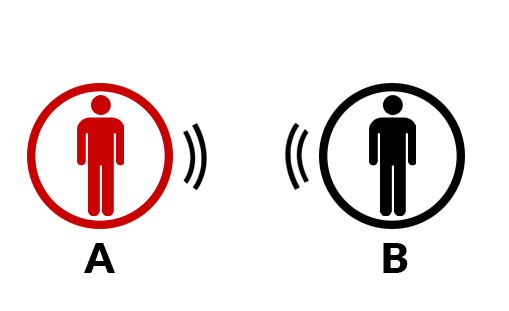}
\caption{Graphical interpretation of patient}
\label{wwf1}
\end{figure}

\begin{figure}
\centering
\includegraphics[scale=1.5]{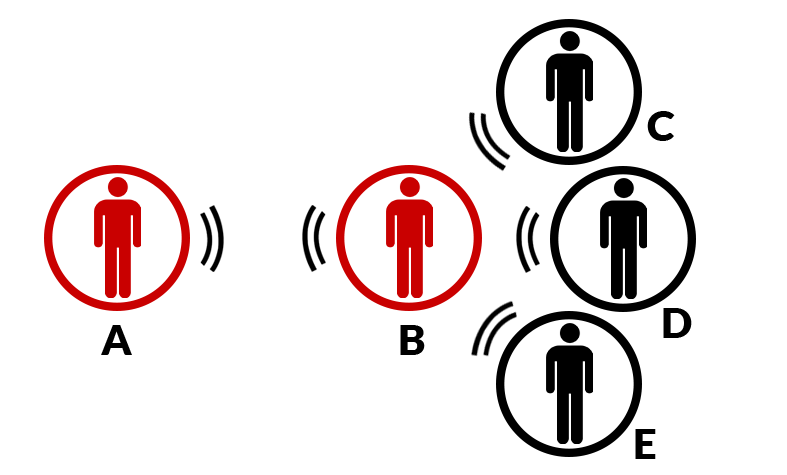}
\caption{Identification of coronavirus patient}
\label{wwf2}
\end{figure}

\begin{figure}
\centering
\includegraphics[scale=1.1]{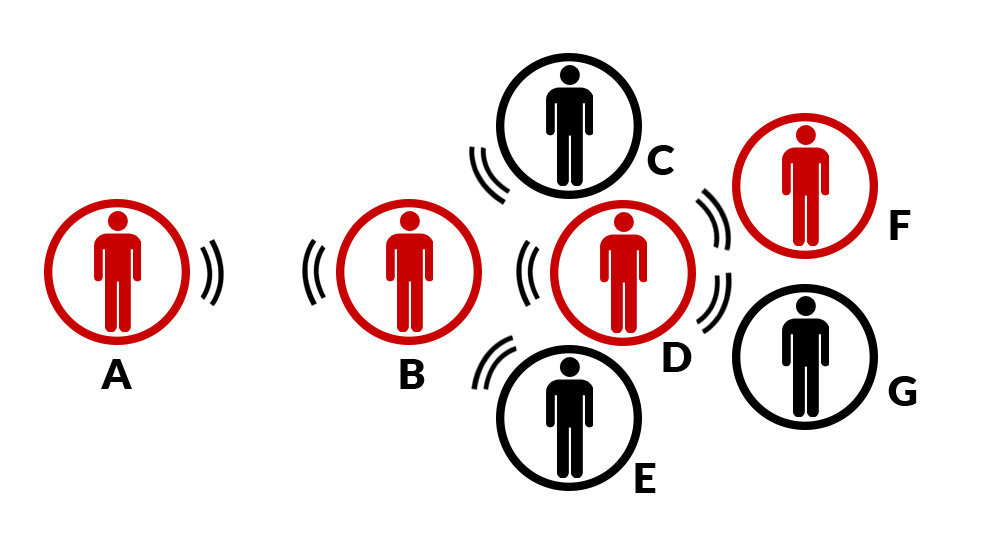}
\caption{Identification of coronavirus patients}
\label{wwf3}
\end{figure}

After few days, person A feels symptoms of COVID-19 and consults a doctor. Unfortunately he is diagnosed and tested positive for COVID-19. Person A opens up his app and verifies his documentation provided by health care provider. He simply taps the button and declares himself COVID-19 patient as shown in Figure \ref{wwf1}. This declaration automatically uploads on storage cloud once Person A declares himself COVID-19 patient. After that Person B phone downloads beacons from all those people who have been diagnosed with COVID-19. In this way Person B can quarantine himself or can perform test for COVID-19. Later if person B diagnoses with COVID-19 he will perform same steps as person A to declare himself COVID-19 patient as shown in Figure \ref{wwf2}. Similarly, if C was in contact with number of people without a phone call then the entire log will be maintained in the same manners as shown in Figure \ref{wwf3}. 

Hence this model is capable to overcome the tracking limitation of CDR when call is not made. This paper presents an efficient and robust framework to track COVID-19 patients based on CDR and ENS. This model will help the governments to track, trace and terminate coronavirus cases by exploring the entire network of COVID-19 patients. 
The capabilities of proposed model is summarized in Table \ref{comparison}.

\begin{table}

\begin{center}
\caption{Capability of proposed model}
\label{comparison}
\begin{tabular}[width=\textwidth]{ | c | c| } 
    \hline
   
    Track and Trace of COVID-19 patients  &   \checkmark  \\
    \hline

   Track and Trace of COVID-19 suspects           &  \checkmark  \\
      \hline
   Monitoring of COVID-19 patients & \checkmark\\
   \hline   
   Path alerts traversed by COVID-19 patients & \checkmark \\
    \hline

\end{tabular}
\end{center}

\end{table}

\section{Conclusion}
\label{conclusion2}
Since there is no vaccine available to cure this pandemic disease, therefore, the best solution is to minimize its spread. This paper has proposed a novel model to control the wide spread of coronavirus using CDRA and contact tracing. The model proposed in this paper can help to track and trace the COVID-19 patients. Through this analysis, one can easily extract the information with whom the patient has remained in contact. So with CDR analysis one can easily track the COVID-19 patients and suspects. Through CDRA, we can easily track the path followed by patients for the last couple of weeks and alerts can be generated for general public to avoid that path. The case study section has thoroughly discussed how to implement this model. One of the main limitations of CDRA is when a suspect goes into the crowd without making a phone call. In order to overcome this limitation contact tracing based on ENS is used in parallel to track the path traversed made by the patient without call history. The information extracted with the help of this hybrid model can easily minimize the wide spread of coronavirus disease. The potential outcomes of the proposed model are tracking of COVID-19 patients, monitoring of isolated patients, tracking of suspected ones and inform the mass about the safest path to use. It is therefore recommended that the proposed model can be adopted by the government to control the wide spread of coronavirus disease. In future, the authors are planning to include different social networks data such as Facebook, WhatsApp, Google, Telegram and Instagram in the proposed model. This will further improve the tracking mechanism of COVID-19 patients.


%




\end{document}